# Near-unity efficiency optical vortex generation in van der Waals materials

Sujeong Byun[1], Munseong Bae[2], Hangsung Cho[3], Haejun Chung[2], Kostya S. Novoselov[4], James Bullock[1], and Sejeong Kim[5,*]

[1]Department of Electrical Engineering, Faculty of Engineering and Information Technology, University of Melbourne, Melbourne 3000, Australia

[2]Department of Electronic Engineering, Hanyang University, Seoul, 04763, Republic of Korea

[3]Department of Nano Engineering, Sungkyunkwan University, Suwon, 16419, Republic of Korea

[4]Department of Materials Science and Engineering, National University of Singapore, 119077, Singapore

[5]Department of Electrical and Computer Engineering, Sungkyunkwan University, Suwon, 16419, Republic of Korea

[*]corresponding author: sejeongk@skku.edu

## Abstract

Optical spin-orbit coupling provides a promising, fabrication-free route for developing u ltra-compact optical vortex generators. However, the conversion efficiency has been theoretically limited to 0.5. Here, we demonstrate enhanced vortex generation efficiency by employing a Bessel beam as the input and propagating it through van der Waals (vdW) crystals. The large birefringence of vdW crystals and the single transverse wave vector of a Bessel beam allow a near unity spin-orbit conversion efficiency and a topological charge transition of $l \rightarrow l + 2$. Through combined analytical and experimental investigations, we demonstrate a conversion efficiency of up to 0.82 in hexagonal boron nitride (hBN) crystals with a thickness of 27.4 μm. The higher efficiency of Bessel input beams over Gaussian beams is attributed to their distinct transverse wave vector distribution of constituent plane wave components. Furthermore, we demonstrate the de pendence of conversion efficiency on the numerical aperture (NA) of the objective lens, which is in good alignment with theoretical predictions. These demonstrations provide a fabrication-free route to highly efficient optical vortex generation via microscale vdW materials platforms.

## 1. Introduction

Optical vortices carrying orbital angular momentum (OAM) [1] have attracted significant interest for applications including high-capacity optical communications [2–5], particle manipulation [6–8], quantum entanglement [9–11], and high-resolution imaging [12, 13]. As such, efficient generation of optical vortex beams has been explored a wide range of approaches including metasurfaces [14–16], spatial light modulators [17–19], and nonlinear processes [20, 21]. However, these techniques typically rely on externally imposed phase modulation, nanofabrication, or require thick (i.e. mm-scale) media for high conversion efficiency.

Spin-orbit coupling (SOC) of light provides an alternative mechanism for manipulating angular momentum by linking the spin state to the spatial phase of an optical field. In anisotropic birefringent media, this effect originates from the differential phase accumulation between two circular polarization states during propagation, enabling deterministic conversion between spin angular momentum and orbital angular momentum. To date, SOC-induced vortex conversion has predominantly relied on Gaussian beam in bulk birefringent crystals [22], typically requiring relatively long propagation distances to accumulate sufficient phase retardation for efficient conversion. Furthermore, the conversion efficiency is fundamentally limited, which prevents complete conversion even at large propagation distances [23]. Thus, reducing the interaction length while maintaining high efficiency remains an important challenge for practical photonic systems implementing optical vortex beams.

In this context, van der Waals (vdW) crystals have recently emerged as a promising platform for SOC-created vortex generation due to their strong optical anisotropy and large intrinsic birefringence [24, 25]. These properties originate from their layered structure, where atoms within the x-y plane are bound by strong covalent bonds, while adjacent layers are held together out of plane by relatively weak van der Waals interactions. This structural anisotropy gives rise to a pronounced difference between the in-plane and out-of-plane refractive indices, as observed in representative vdW materials such as hexagonal boron nitride (hBN) [26], molybdenum diselenide [27], which both exhibit significant birefringence across the visible range. This characteristic enables phase accumulation over shorter propagation lengths compared to conventional bulk crystals, allowing on-chip implementation. Previous demonstrations of optical vortex generation using vdW materials have been conducted with a Gaussian input beam, which has an intrinsic efficiency limit of 0.5 [28]. The use of structured light, in combination with vdW materials, to overcome this intrinsic efficiency limit has yet to be experimentally proven.

Bessel beams are a prominent example of structured lights that have been widely studied because of their non-diffracting nature and their unique angular spectrum [29–31]. These properties have been utilized for optical vortex generation and angular momentum control [15, 32–35]. However, these approaches are primarily limited by the relatively long propagation distances required to generate efficient optical vortex beams or by the reliance on external electric control. Consequently, achieving deterministic and near-unity spin-orbit conversion in compact and strongly birefringent platforms still remains a challenge.

Here, we demonstrate highly efficient optical vortex generation in a vdW crystal enabled by a Bessel input beam. We first present the spin-orbit-induced optical vortex generation and its evolution with respect to the propagation distance, showing a deterministic topological charge transition of $l \rightarrow l+2$. We then investigate theoretically and experimentally how the conversion behavior varies with crystal thickness, from 18.5 μm to 50 μm, and numerical aperture by tuning these parameters. The experimental results agree well with the analytical model, showing that the conversion efficiency approaches unity under appropriate thicknesses and numerical apertures, surpassing the theoretical limit of the Gaussian incident beam case. To elucidate the underlying physics intuitively, we provide a momentum-space interpretation based on the k-sphere representation, where the constant transverse wave vector characteristic of a Bessel beam suppresses dephasing compared to the broad transverse wave vector distribution of a Gaussian beam. By combining the single transverse wave vector of the Bessel beam with the strong birefringence of vdW crystals, we demonstrate high-purity and near-unity spin-orbit conversion within a compact propagation distance.

## 2. Results and Discussion

We consider light propagation in a uniaxial crystal with an optical axis aligned along the z-direction.

In this uniaxial crystal with dielectric tensor

$$\varepsilon = \begin{bmatrix} n_o^2 & 0 & 0 \\ 0 & n_o^2 & 0 \\ 0 & 0 & n_e^2 \end{bmatrix} \tag{1}$$

where $n_o$ and $n_e$ are the ordinary and extraordinary refractive indices, respectively, each plane wave component splits into ordinary and extraordinary components.

An arbitrary field propagating along the $z$-direction can be expressed through its angular spectrum representation as

$$E(\mathbf{r}_\perp,z) = \iint d^2\mathbf{k}_\perp [\tilde{A}_o(\mathbf{k}_\perp,0)\,\mathbf{u}_o(\mathbf{k}_\perp)\,e^{ik_{oz}z} + \tilde{A}_e(\mathbf{k}_\perp,0)\,\mathbf{u}_e(\mathbf{k}_\perp)\,e^{ik_{ez}z}]\,e^{i\mathbf{k}_\perp \mathbf{r}_\perp} \tag{2}$$

Here, $\mathbf{r}_\perp = (x,y)$ denotes the transverse spatial coordinate, $\mathbf{k}_\perp = (k_x,k_y)$ represents the transverse components of the wavevector; $\tilde{A}_{o,e}(\mathbf{k}_\perp,z)$and $\mathbf{u}_{o,e}(\mathbf{k}_\perp)$describe the angular spectrum amplitude at the propagation distance z and polarisation vectors of o- and e- wave polarization components of each plane wave, respectively. Also, $k_{oz} = \sqrt{k_0^2 n_o^2 - k_\perp^2}$ and $k_{ez} = \sqrt{k_0^2 - \frac{k_\perp^2}{n_e^2}}$ are wave vector components of o- and e-waves in the z-direction obtained by the dispersion relation, where $k_0 = 2\pi/\lambda$ is the vacuum wavenumber with $\lambda$ being the wavelength. $\mathbf{u}_{o,e}(\mathbf{k}_\perp)$ can be expressed as

$$\begin{aligned} \mathbf{u}_o(\mathbf{k}_\perp) &= -\sin\phi\hat{\mathbf{x}} + \cos\phi\hat{\mathbf{y}}, \\ \mathbf{u}_e(\mathbf{k}_\perp) &= \frac{k_{ez}}{n_o^2 k_e}\frac{\mathbf{k}_\perp}{k_\perp} - \frac{k_\perp}{n_e^2 k_e}\hat{\mathbf{z}}, \end{aligned} \tag{3}$$

where $\phi$ is the azimuthal angle of plane wave components in the spherical coordinate system, and $k_{o,e} = n_{o,e}k_0$. Now we consider the circular polarisation (CP) basis and the subscripts + and - represent left- and right-handed circular polarisation, respectively. When we apply the incident beam with cylindrical symmetry at z = 0 position, Eq. 2 can be written as follows:

$$E(\mathbf{r}_\perp,0) = \iint d^2\mathbf{k}_\perp e^{i\mathbf{k}_\perp \mathbf{r}_\perp} \left[\tilde{U}_+(\mathbf{k}_\perp,0)\,\widehat{\mathbf{V}}_+ + \tilde{U}_-(\mathbf{k}_\perp,0)\,\widehat{\mathbf{V}}_-\right], \tag{4}$$

where $\tilde{U}_\pm(\mathbf{k}_\perp,0)$ are the two-dimensional Fourier transform of the transverse field at $z = 0$ of each handedness, and $\widehat{\mathbf{V}}_\pm \equiv (\hat{\mathbf{x}} \pm i\hat{\mathbf{y}})/\sqrt{2}$ are unit vectors of each circular polarisation. Let us assume that the incident beam has left-handed circular polarisation. In this case, the field at distance $z$ can be written as

$$\begin{aligned} \tilde{U}_+(\mathbf{k}_\perp,z) &= t_{++}(\mathbf{k}_\perp,z)\,\tilde{U}_+(\mathbf{k}_\perp,0), \\ \tilde{U}_-(\mathbf{k}_\perp,z) &= t_{-+}(\mathbf{k}_\perp,z)\exp(i2\phi)\tilde{U}_+(\mathbf{k}_\perp,0), \end{aligned} \tag{5}$$

where $t_{++} = \left[\exp\left(ik_{ez}z\right) + \exp\left(ik_{oz}z\right)\right]/2$ and $t_{-+} = \left[\exp\left(ik_{ez}z\right) - \exp\left(ik_{oz}z\right)\right]/2$, denoting the transmission amplitudes of the non-spin-flipped and spin-flipped mode, respectively.

Equation 5 describes the evolution of the circularly polarised field components in the transverse wave vector representation during the propagation through the uniaxial crystal. As the ordinary and extraordinary eigenmodes travel inside the material, they accumulate different propagation phases, $\exp\left(ik_{oz}z\right)$ and $\exp\left(ik_{ez}z\right)$, determined by their respective longitudinal wave vectors $k_{oz}$ and $k_{ez}$. The resulting phase difference between these two modes leads to the interference between the ordinary and extraordinary contributions. Importantly, the spin-flipped component acquires an additional azimuthal phase factor $\exp(i2\phi)$, indicating that the converted field carries orbital angular momentum with a topological charge of +2. This shows that the spin-orbit coupling is induced by the birefringence of the crystal.

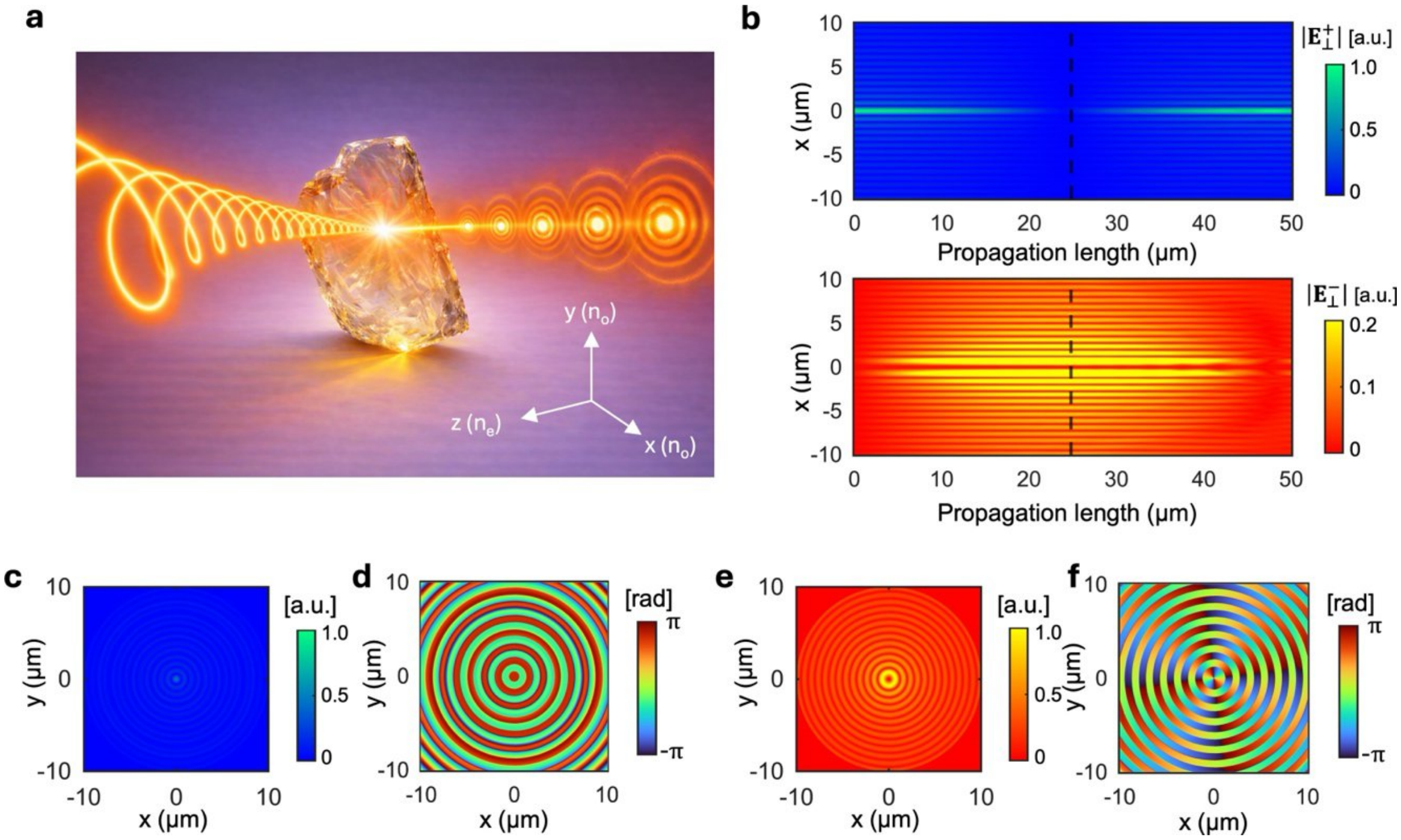

Figure 1: Optical vortex generation within a hBN crystal with a Bessel incident beam. (a) Schematic illustration of optical vortex generation with a Bessel beam. A Bessel beam with circular polarisation on the right hand side propagates through the hBN crystal along with the optic axis i.e. extraordinary axis. As a consequence, part of the entire transmitted beam shows the opposite handedness and twisted wavefront, imparting the topological charge of ±2 depending on the handedness of the incident polarisation. (b) Simulated amplitudes of left-handed circularly polarised (LCP, $|\mathbf{E}_+|$, top) and right-handed circularly polarised (RCP, $|\mathbf{E}_-|$, bottom) electric field with respect to the propagation length within the hBN crystal is obtained with LCP Bessel input beam. The black dashed line indicates $z = 23$ μm. They are normalised by the peak amplitude of LCP electric field. (c) Intensity profile of the LCP component at $z = 23$ μm. (d) Corresponding phase profile. (e) Intensity profile of the RCP component at the same propagation distance as (c). (f) Corresponding phase profile. The spiraling arms at the centre confirm the generation of an optical vortex with topological charge of +2.

To confirm this theoretical description above, we perform numerical simulations of beam propagation through a hBN crystal. The configuration is illustrated schematically in Fig. 1a, where a circularly polarised Bessel beam propagates along the optic axis of the crystal. The birefringence leads to a relative phase accumulation between the ordinary and extraordinary eigenmodes, which results in the generation of a spin-flipped component carrying an azimuthal phase factor $\exp(i2\phi)$. The simulated evolution of the circularly polarised field components with respect to the propagation distance within the hBN crystal with the wavelength of 594 nm and the numerical aperture of 0.4 is shown in Fig. 1b for LCP Bessel input beam. The refractive indices of the hBN crystal is taken from [26], which is $n_o = 2.15$ and $n_e = 1.86$ at $\lambda = 594$ nm. The simulation results with LCP Gaussian input beam is provided in Fig.

S1a and b, for LCP and RCP field, respectively.

For both cases, the initially incident LCP field gradually transfers energy to the RCP component as the beam propagates through the crystal. However, the spatial evolution of the converted component differs markedly for the two types of incident beams. While the Gaussian beam exhibits a gradual growth of the RCP field along the propagation direction, the Bessel beam produces a more pronounced conversion, leading to a stronger RCP component over the same propagation distance. In particular, the electric field of RCP component reaches the maximum at $z = 23$ μm.

To further examine the resulting field structure, the intensity distributions at $z = 23$ μm in $xy$ plane are presented in Fig. 1c and e for the LCP and RCP components, respectively, for the Bessel beam case. Intensity distributions with the Gaussian incident beam are given in Fig. S1c and e for LCP and RCP fields, respectively. The converted RCP component exhibits a characteristic intensity profile with a phase singularity at the centre for both cases, indicating the presence of an optical vortex. The corresponding phase distributions are shown in Fig. 1d and f, where the RCP component clearly displays a helical phase structure with a phase shift of $4\pi$. This phase singularity confirms the generation of an optical vortex with a topological charge of +2, consistent with the theoretical prediction of the spin-orbit coupling mechanism. The Gaussian beam case shows similar results in Fig. S1d and f.

We construct a transmission setup using a hBN crystal as a sample to experimentally verify the optical vortex generation shown by the simulations. The experimental configuration is depicted in Fig 2a. A laser beam with a wavelength of 594 nm is first converted to circularly polarised light using a linear polariser and a quarter waveplate. The beam is then focused onto the hBN crystal on the glass substrate, where the spin-flipped components carrying orbital angular momentum is generated owing to spin-orbit coupling. The transmitted beam is analysed using a quarter waveplate and a linear polariser, allowing the intensity distribution and interference patterns of the LCP and RCP components to be measured separately.

In Figure. 2b and c, the obtained intensity profiles for non-spin-flipped i.e. LCP and spin-flipped i.e RCP components with or without reference beams are shown. Intensity profiles for both Gaussian and Bessel input beams, with a hBN crystal thickness of 27.4 μm, are provided. Figure. 2b shows the measured intensity profiles only with the hBN crystal on the glass substrate for a Gaussian and a Bessel incident beam. In both cases, the converted RCP component exhibits a doughnut-shaped intensity profile, which is the characteristic of an optical vortex.

To further confirm the topological charge of orbital angular momentum, interference measurements are done by introducing a reference beam path. The interference patterns are presented in Fig. 2c. The spiral arms observed in the RCP component profiles indicate a helical phase structure, providing the topological charge of an optical vortex of +2 in both cases.

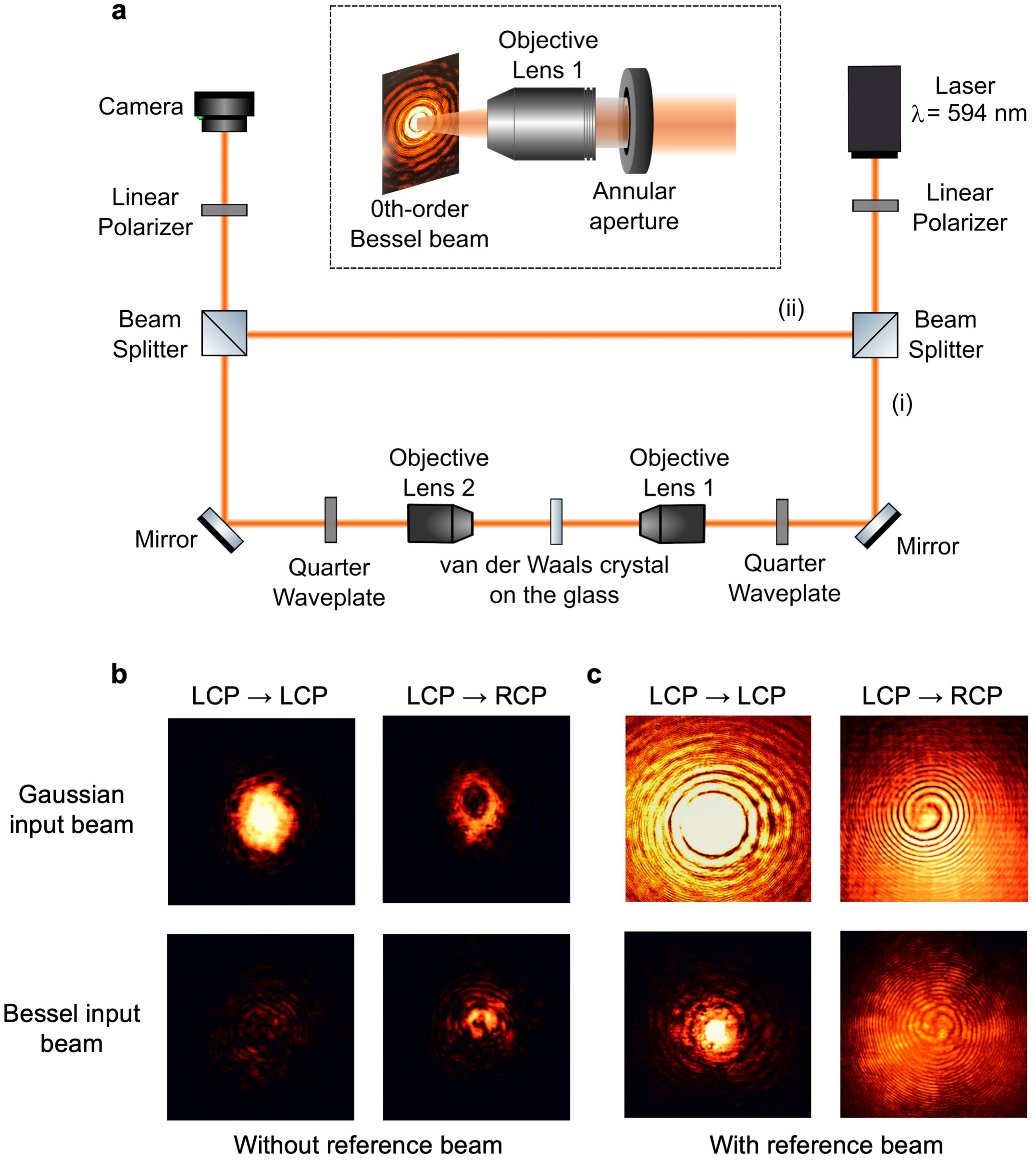

Figure 2: Demonstration of optical vortex generation using a hBN crystal. (a) The transmission setup used for the optical vortex generation with either a Gaussian beam or a Bessel input beam. An input beam with the wavelength of 594 nm directed along the beam path (i) is converted into left-or right- handed circularly polarised light as it goes past a linear polariser and a quarter waveplate. The circularly polarised light propagates within the vdW crystal on the glass resulting in an output beam composed of both left and right- handed circularly polarised light. A quarter waveplate and a linear polariser placed afterwards enable the observation of the intensity profile of generated circularly polarised light only. The reference beam path (ii) is added to obtain the interference pattern of the optical vortex. The inset shows the method to make 0th-order Bessel beam by placing the annular aperture between a quarter waveplate and the objective lens 1. (b) Obtained intensity profiles using the hBN crystal with a thickness of 27.4 μm without reference beam with a Gaussian (top) and a Bessel (bottom) input beam. The input beam has a left-handed circular polarisation. the non-spin-flipped (LCP) and spin-flipped (RCP) components are shown on left and right column, respectively. (c) Corresponding interference patterns obtained by adding the reference beam.

To quantify the spin-orbit coupling, the conversion efficiency is defined as the ratio of the power of the spin-flipped component to the total input power as described below [36]:

$$\eta(z) = \frac{\iint \left|\tilde{U}_-(\mathbf{k}_\perp,z)\right|^2 d^2\mathbf{k}_\perp}{\iint \left|\tilde{U}_+(\mathbf{k}_\perp,0)\right|^2 d^2\mathbf{k}_\perp}. \tag{6}$$

This equation provides a general definition of the conversion efficiency in the angular spectrum representation. Therefore, the resulting conversion efficiency depends on the spectral distribution of the input beam. For a LCP Gaussian beam, the angular spectrum at $z = 0$ is given by

$$\tilde{U}_+(\mathbf{k}_\perp,0) = \frac{\omega_0^2}{2}\exp\left(-\frac{\omega_0^2 k_\perp}{4}\right), \tag{7}$$

where $\omega_0$ denotes the beam waist. Substituting this spectrum into Eq. 6 and evaluating the integral under the paraxial approximation yields the following expression for the conversion efficiency [23]:

$$\eta_G = \frac{1}{2}\left[1 - \frac{1}{1 + \left(\frac{z}{L}\right)^2}\right], \tag{8}$$

where $L = k_0 n_o \omega_0^2 / \left(n_0^2/n_e^2 - 1\right)$. This result shows that conversion efficiency of a Gaussian beam asymptotically approaches 0.5 as the propagation distance increases.

On the other hand, an ideal Bessel beam is composed of plane-wave components with a fixed transverse wave vector magnitude. Accordingly, its angular spectrum is concentrated on a narrow ring in *k*-space and can be expressed as

$$\tilde{U}_{+}(k_{\perp},0) = \tilde{U}_0 \delta(k_{\perp} - \Delta k), \tag{9}$$

where $\tilde{U}_0$ denotes the arbitrary amplitude. Substituting this spectrum into Eq.6 leads to conversion efficiency

$$\eta_B = \frac{1}{4} \left|\exp(ik_{ez}z) - \exp(ik_{oz}z)\right|^2 = \sin^2\left(\frac{k_{oz} - k_{ez}}{2} z\right). \tag{10}$$

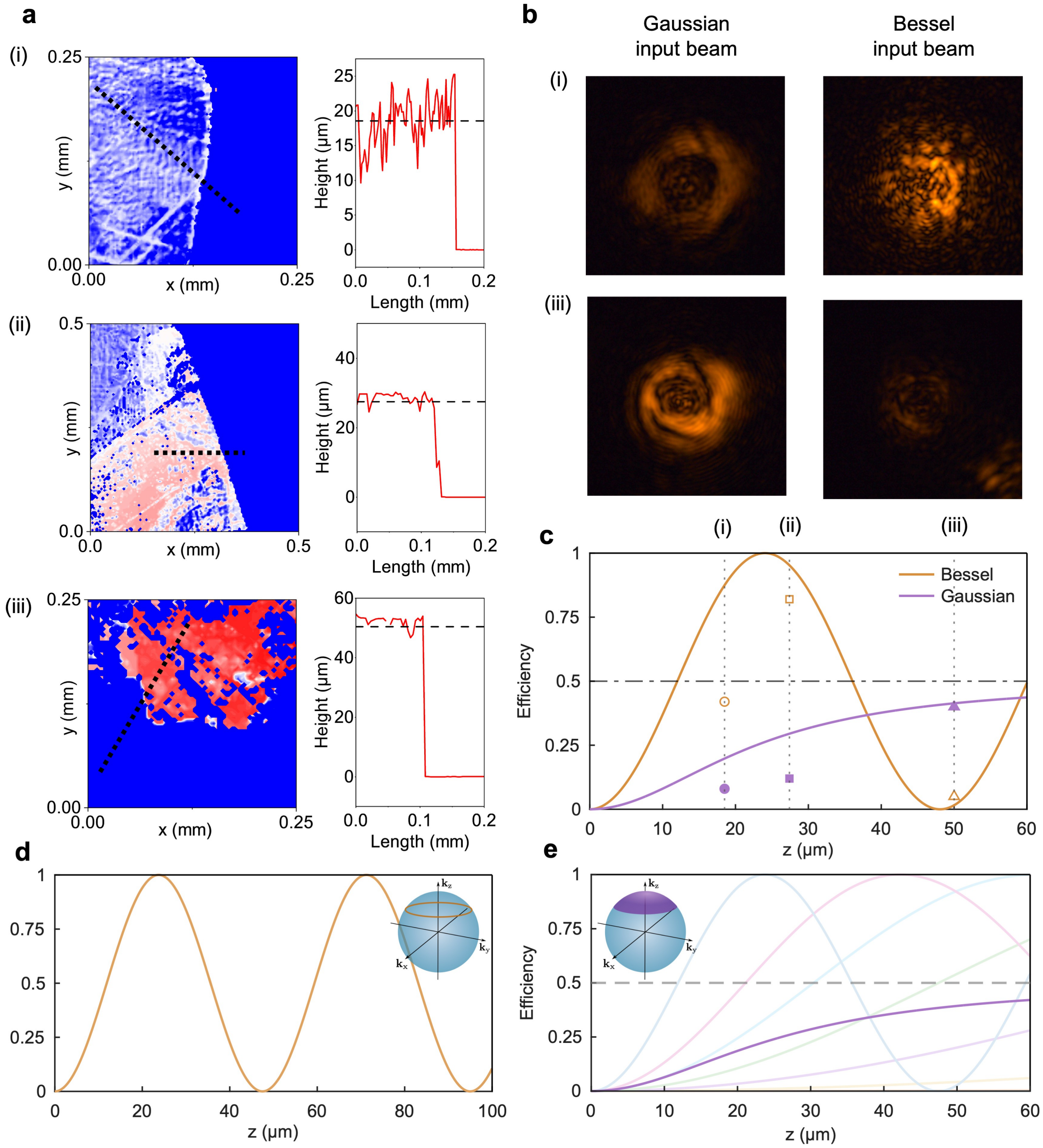

Figure 3: Thickness-dependent optical vortex generation. (a) Thickness characterisation of hBN crystals. (left) Optical profilometry height maps. (right) Cross-sectional height profiles along black dotted lines in height maps. Average thicknesses are shown as black dashed lines; (i) 18.5 μm, (ii) 24 μm, and (iii) 50 μm. (b) Measured intensity profiles for left-handed circularly polarised Gaussian (left) and Bessel (right) input beams. Samples correspond to thicknesses (i) 18.5 μm and (iii) 50 μm. (c) Con-version efficiency as a function of propagation distance z. Solid lines show simulated results for Bessel (ochre) and Gaussian (purple) input beams, whereas markers represent experimental measurement results. The horizontal dashed line indicates efficiency of 0.5, and (i)-(iii) labels identify different propagation distances. (d) Conversion efficiency for a Bessel beam characterised by a constant transverse wave vector. The inset illustration is the transverse wave vector of a Bessel beam on the k-sphere, rep-resented by the ring shape on the surface. (e) Conversion efficiency for a Gaussian beam decomposed into multiple transverse wave vector components. The faint curves represent the conversion efficiency of individual transverse wave vector components, whereas the purple line denotes the efficiency integrated over all transverse wave vectors. The inset image shows that transverse wave vectors of the Gaussian beam cover a broad region of the surface of the k-sphere near the central wave vector.

Unlike the Gaussian case, the efficiency of the Bessel beam can reach unity at a specific propagation distance, whereas the efficiency of the Gaussian beam remains bounded below unity. The origin of this difference will be discussed later by examining the conversion process.

As described in Eq. 8 and 10, the spin-orbit conversion efficiency is proportional to the propagation distance within the uniaxial crystal. In this regard, we obtain conversion efficiencies both analytically and experimentally using hBN crystals with various thicknesses. Fig. 3a shows the thickness information of selected hBN samples measured by the optical profilometer. To span the wide range of propagation distances, three hBN crystals with the thickness of 18.5 μm, 27.4 μm, and 50 μm are selected. The corresponding intensity profiles of the converted component for Gaussian and Bessel incident beams for samples with thicknesses of (i) 18.5 μm and (iii) 50 μm are shown in Fig. 3b. Note that intensity profiles of sample (ii) are given in Fig. 2b and c. The intensity profile is relatively clear with the Gaussian incident beam, whereas the intensity distribution in the Bessel input beam case shows more irregular and distorted intensity profile. This behaviour can be attributed to the different responses to imperfections in the crystal, such as thickness variations and surface roughness [37]. Nevertheless, the relative intensity of the converted component can still be clearly identified. In particular, the Bessel beam case in sample (i) produces a stronger converted signal, although the intensity profile with the Gaussian beam is brighter in sample (iii). This trend is quantitatively confirmed in Fig. 3c.

Solid lines in Fig. 3c denote the analytically calculated conversion efficiencies with Gaussian and Bessel beams depending on the propagation distance obtained by Eq. 8 and 10, respectively. In the Gaussian beam case, it gradually approaches 0.5 at maximum as the beam travels within the material. On the other hand, the conversion efficiency with the Bessel beam oscillates, reaching unity when the propagation distance is 23 μm. The markers represent the experimentally obtained values for the three different thicknesses, using the same method as our previous study [28], showing a good agreement with the analytically calculated conversion efficiencies. While the ideal Bessel beam supports unity conversion, experimental demonstration are affected by finite aperture size and crystal inhomogeneities, resulting in slightly lower values.

Now, we investigate the discrepancy of conversion efficiencies between a Gaussian and Bessel beam case in transverse wave vector domain. First, the conversion efficiency for each transverse wave vector channel can be defined using Eq.5, written as

$$\eta_{\mathbf{k}_\perp}(z) \equiv \frac{\left|\tilde{U}_-(\mathbf{k}_\perp,z)\right|^2}{\left|\tilde{U}_+(\mathbf{k}_\perp,0)\right|^2} = |t_{-+}(\mathbf{k}_\perp,z)|^2 = \sin^2\left(\frac{k_{oz}-k_{ez}}{2}z\right). \tag{11}$$

Due to the sinusoidal waveform, this indicates that the individual $\mathbf{k}_\perp$ channel can in principle experience the complete conversion to spin-flipped state with an appropriate propagation distance. However, the real beam generally consists of a distribution of transverse wave vectors described by its respective angular spectrum. Therefore, the realistic conversion efficiency of the beam should account for the contributions of all the transverse wave vector channel. Since each channel supplies the beam with a weight proportional to its intensity $\left|\tilde{U}_+(\mathbf{k}_\perp,0)\right|^2$, the overall conversion efficiency is obtained by taking the intensity-weighted average of $\eta_{\mathbf{k}_\perp}(z)$ over the angular spectrum of the input field,

$$\eta(z) = \frac{\iint \eta_{\mathbf{k}_\perp}(z)\left|\tilde{U}_-(\mathbf{k}_\perp,z)\right|^2 d^2\mathbf{k}_\perp}{\iint \left|\tilde{U}_+(\mathbf{k}_\perp,0)\right|^2 d^2\mathbf{k}_\perp}. \tag{12}$$

Since $\tilde{U}_+(\mathbf{k}_\perp,0)$ is localised on a narrow annular shell at $k_\perp = k_{0\perp}$ for ideal Bessel beam as shown in the inset image of Fig. 3d, this equation reduces to the expression for the single channel,

$$\eta_B(z) = \sin^2\left[\frac{\left(k_{ez}(k_{0\perp}) - k_{oz}(k_{0\perp})\right)}{2}z\right]. \tag{13}$$

This result is the same as Eq. 10, which can reach unity at the appropriate thickness as well. The behaviour of the conversion efficiency with a Bessel beam according to this equation is plotted when $k_{0\perp} = 0.4k_0$ in Fig. 3d.

In contrast, a Gaussian beam is composed of plane waves with various $k_\perp$ as depicted in the inset k-sphere image of Fig. 3e and the angular spectrum of each channel is described as Eq. 7. As the beam propagates within the crystal, each $k_\perp$ channel acquires a different phase due to their varied spatial period, thereby leading to dephasing across the angular spectrum. Consequently, the oscillating behaviour observed for a single plane wave disappears after averaging over all transverse wave vectors. As a result, the phase difference among channels becomes random at large propagation distances, causing the contributions from each channel to average out to their mean value, 0.5. The conversion efficiency obtained by averaging over multiple $k_\perp$ channels can be expressed by substituting Eq. 7 into Eq. 12 as

$$\eta_G(z) = \frac{\int_0^\infty \eta_{k_\perp}(z) \exp\left(-\frac{\omega_0^2 k_\perp^2}{2}\right) k_\perp dk_\perp,}{\int_0^\infty \exp\left(-\frac{\omega_0^2 k_\perp^2}{2}\right) k_\perp dk_\perp}. \quad (14)$$

The calculated conversion efficiencies of individual $k_\perp$ channel and overall conversion efficiency i.e. intensity-weighted average are shown in Fig. 3e, confirming the theoretical bound of approximately 0.5.

The NA corresponds to the transverse wave vector of a Bessel beam, given by $k_\perp = NA \times k_0$. Fig. 4a shows analytically calculated conversion efficiencies as a function of propagation distance for different NA values. The result indicates that the efficiency can reach unity at specific propagation distances, and that increasing NA shifts the required propagation distance needed to reach unity. This behaviour arises because larger transverse wave vector leads to a greater difference between longitudinal wave vectors of ordinary and extraordinary components, resulting in a faster conversion process along the propagation direction. To experimentally verify this phenomenon, the conversion efficiencies are measured as a function of NA for crystals with different thicknesses, as shown in Fig. 4b. The solid curves represent the analytically calculated conversion efficiencies at fixed propagation distances corresponding to each sample thickness, while the markers denote the experimentally obtained values for samples with thicknesses of (i) 18.5 µm, (ii) 27.4 µm, and (iii) 50 µm. The mismatch for the 18.5 µm crystal is presumably associated with uncertainty in the profilometry measurement. Because the surface step exhibited considerable noise as shown (i) in Fig. 3a, the actual thickness may differ slightly from the nominal value. For each thickness, the measured efficiencies follow the corresponding calculated values, with the maximum efficiency at different NA values depending on the thickness. In particular, thinner samples achieve unity at large NA relative to thicker samples. Increasing the NA enhances the transverse wave vector, thereby increasing the phase difference between the ordinary and extraordinary waves, which accelerates the spin-orbit conversion process. The overall agreement between experiment and calculation confirms that the conversion efficiency is governed by the NA-dependent propagation dynamics of the Bessel beam.

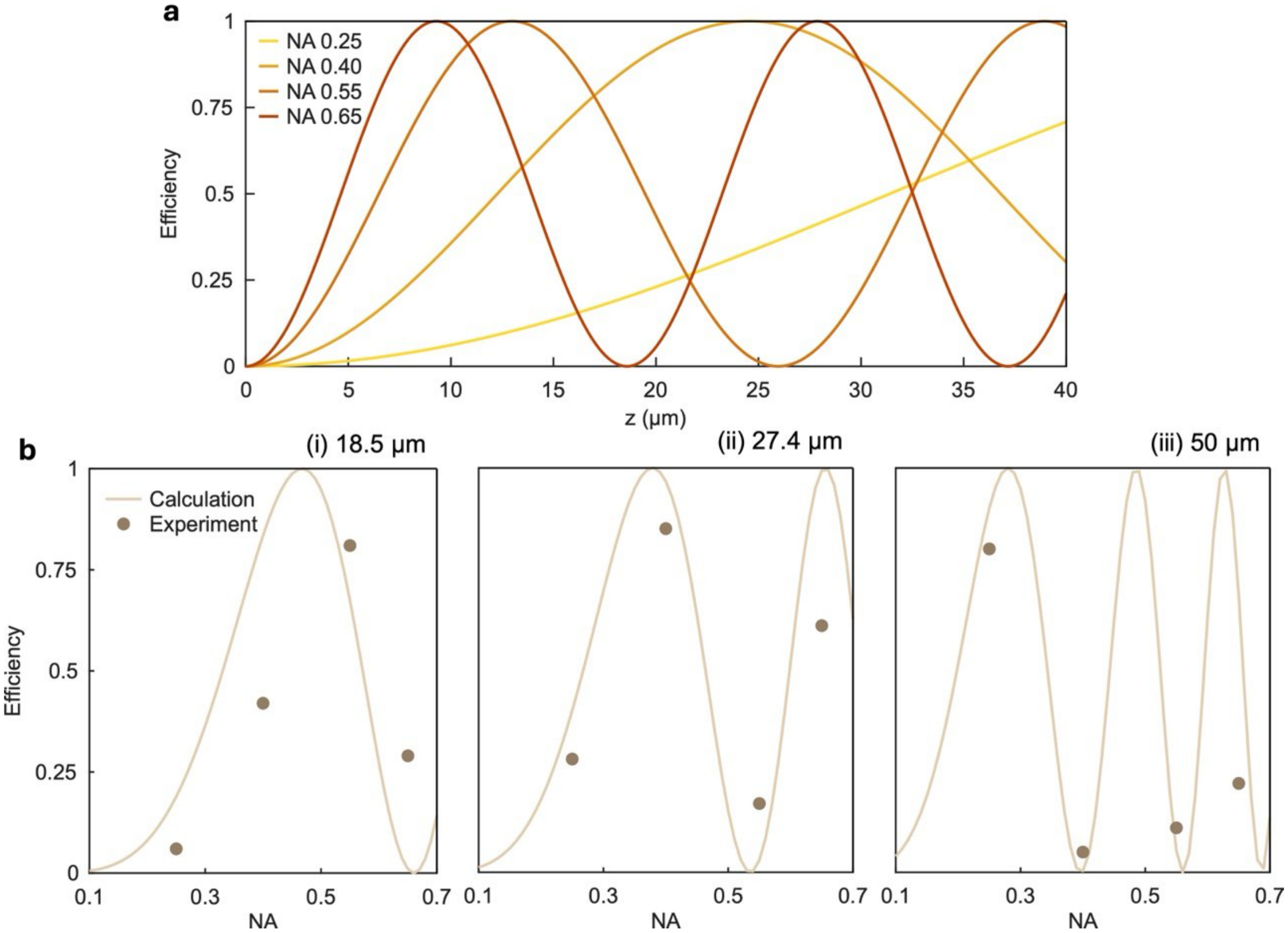

Figure 4: Numerical aperture-dependent conversion efficiency. (a) Analytically calculated conversion efficiencies for a Bessel beam as a function of propagation length for different NA values. Larger NA leads to a shorter propagation distance required to reach unity efficiency. (b) Comparison between the calculated and experimentally measured conversion efficiencies as a function of NA values for different thicknesses of (i) 18.5 μm, (ii) 27.4 μm, and (iii) 50 μm. Solid curves represent the analytical calculations, while the dots denote the experimental results, using the NA values outlined in (a).

## 3. Conclusion

In this work, we have demonstrated highly efficient optical vortex generation in a van der Waals (vdW) crystal enabled by a Bessel incident beam. By applying spin-orbit coupling in a hBN crystal, we showed that a circularly polarised Bessel beam undergoes the conversion into an optical vortex with a topological charge of $l \rightarrow l+2$, accompanied with the efficient spin-flipping. Through both analytical calculation and experimental measurements, we have shown that the conversion efficiency strongly depends on the transverse wave vector distribution of the incident beam. As a result, near-unity spin-orbit conversion can be achieved at specific propagation distances. Experimentally, using the hBN crystal with the thickness of 27.4 μm, we observe good agreement with theory, achieving a conversion efficiency of approximately 0.82. This value exceeds the efficiency obtainable with a Gaussian beam under the same conditions, with ~7-fold higher efficiency. We further examined the thickness-dependence by exploring hBN crystals with thicknesses of 18.5 μm and 50 μm, showing agreement with analytically anticipated efficiencies. An explanation of the origin of the efficiency enhancement with a Bessel beam using the transverse wave vector domain has been given. Further, the dependence of efficiency on NA is shown, with all three thicknesses achieving a conversion efficiency above 0.8, at specific NA values. This behaviour arises due to the increase of the transverse wave vector, following the increase of NA. We believe that this study can provide a straightforward understanding of why a Bessel beam enables higher spin-orbit conversion efficiency.

## 4. Acknowledgment

The thickness measurement was performed in part at the Melbourne Centre for Nanofabrication (MCN) in the Victorian Node of the Australian National Fabrication Facility (ANFF). This paper was supported by Korea Institute for Advancement of Technology (KIAT) grant funded by the Korea Government(MOTIE)(RS-2024-00418086,HRDProgramforIndustrialInnovation).